# Elucidating the Size of Chemical Space with Assembly Theory

Juan Carlos Morales Parra[1†], Keith Y Patarroyo[1†], Abhishek Sharma[1], David Obeh Alobo[1] and Leroy Cronin[1,2]*
[1]School of Chemistry, The University of Glasgow, University Avenue, Glasgow G12 8QQ, UK
[2]Santa Fe Institute, Santa Fe NM, USA

*Corresponding author email: Lee.Cronin@glasgow.ac.uk [†]Equal contribution.

**Abstract:** Chemical space is unimaginably vast with common heuristic estimates suggesting that there are *ca.* $10^{60}$ 'drug-like' molecules possible below a molecular mass of 500 Da. However, these estimates largely ignore the structural and synthetic complexity of the molecules enumerated. Here we present a first-principles estimate of the size of chemical space using the Assembly Theory, which quantifies the amount of causation required to form a molecule, captured in the assembly Index. This is a measurable molecular complexity measure derived from the minimum number of recursive bond-joining operations required to construct a molecular graph. Assembly Theory partitions chemical space into levels defined by Assembly Index, allowing bounds to be placed on its growth as molecular complexity increases. We show that chemical space (the accumulated Assembly Index level sets) grows at least super-exponentially, and at most, double-exponentially with respect to the Assembly Index. Using the GDB-13 database as a reference for growth-rate estimation, we model how chemical space expands under increasing complexity and contracts under structural constraints, including atom and bond types, number of rings, ring size, and chemical motifs. Under constraints comparable to standard drug-like estimates, including molecular mass below 500 Da, our analysis yields a chemical space of approximately $10^{117}$ molecules at Assembly Index 25. Finally, we constrain chemical space by biologically relevant motifs and identify structurally relevant molecules near the accessible boundaries of these assembly-defined spaces.

## Introduction

The nature of chemical space is of fundamental importance for chemistry, medicine[1] and material discovery[2]. Features such as the sparsity of useful molecules in this vast space are, in part, a consequence of the combinatorial nature of chemical space[3]. Therefore, quantifying the fundamental constraints on this space is one tool for navigating this landscape, and it may yield insights into not only drug and material design but also into the nature of other combinatorial spaces such as genetic and protein spaces. Given a set of potential energetic or structural constraints, it is useful to estimate the size of the space of molecules that satisfy those constraints (e.g. including only a selected set of structures, types of bonds or exhibiting fixed topological structures like rings or containing a fixed motif) since this indirectly quantifies how much the introduction of these restrictions shrinks the space of all possible combinatorial molecules. Different applications, such as drug discovery[4] involve the exploration of constrained regions of chemical space and quantifying the relative size of these with respect to the full combinatorial space is an indicator of how difficult is the task of targeting a specific molecule located in a given constrained space.

Chemical space can be defined explicitly or implicitly[4] depending on the resources available. Explicit approaches rely on chemical libraries of explored[5] or enumerated[6] compounds. Although these libraries are vast, computational limitations restrict the portion of the space that can be fully explored. Implicit approaches overcome this limitation by defining the chemical space available to explore on the fly; they generate molecules based on an available pool of building blocks and rules on how to grow the desired structure. These methods are used in genetic algorithms[7–9] or in current deep learning[10–12] frameworks to generate molecular structures. Although these methods can generate large and interesting structures, the portion of the space that is being sampled is harder to quantify than with the explicit methods.

Here we take an explicit approach to study the size and growth of chemical space. Previous approaches to studying the size of chemical space have relied on estimates of the total number of possible organic molecules that can be combinatorially devised[1]. Other approaches have relied on power-law extrapolation[13] or on enumerating all possible organic molecules up to a number of atoms and using filters to find useful drug candidates[6,14–16]. These enumerated databases or generated databases (GDBs) have also been studied from a molecular complexity point of view[17]. More recently, Assembly Theory[18] has emerged as a framework for the study of recursive construction of discrete objects and their causal complexity. It has been used to explore organic chemistry[19], biochemical space[20], crystalline material complexity[21] and large-scale similarity exploration[22] in chemical space. Assembly Theory provides a molecular complexity (the Assembly Index of the molecule) that better captures duplicate symmetry[23] than other complexity metrics and is distinct from other information measures[24].

In this paper, we explore the nature of chemical space from an Assembly Theory perspective and answer the question of how many molecules are possible at a given level of *complexity*, Figure 1a. The Assembly Index of a molecule is defined as the minimum number of recursive bond-joining operations required to build the molecule. This intrinsic quantity allows us to partition any chemical space into levels labelled by the Assembly Index of the molecules in it. If $a>b$ then the molecules with Assembly Index *a require more recursive construction steps than molecules labelled by b*; so implicitly this partition also defines a *constructive ordering* in the chemical space. Hence by tracking the sizes of the level sets for increasing Assembly Indices we quantify the *growth* of the chemical space in terms of the constructability of the molecules, Figure 1b. Even though other quantities could also be considered in order for labelling, partitioning and ordering the chemical space (e.g. size of the molecule (number of bonds), molecular mass or any molecular complexity measure), the choice of the Assembly

Index provides an explicit and unique interpretation of the growth in terms of the minimal number of recursive causal steps required to construct the molecules.

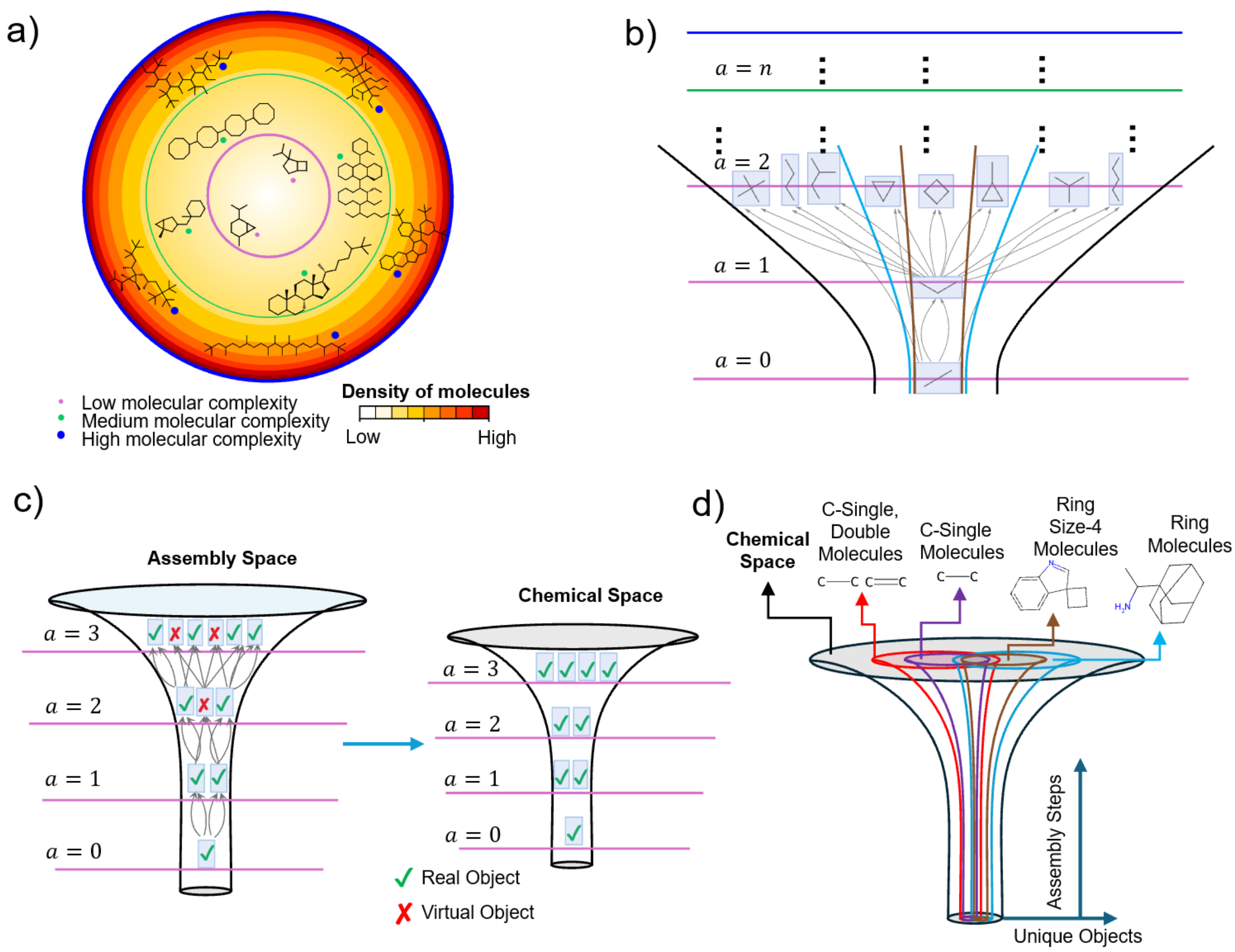

**Figure 1. Growth of chemical space.** (a) Combinatorial explosion of chemical space. The number of molecules with a given molecular complexity exhibits an explosive growth, represented by the radius of the sphere. (b) Expanding molecular space is represented as the assembly space of all molecules. This space is indexed by the Assembly Index. Note that the space of molecules with rings(blue) and the space of molecules with rings of size four(brown) grow at a slower rate. (c) The assembly space is defined by construction pathways whose elements are observed object types, fragments, and virtual objects. For size of chemical space analysis, we only consider observed object types, but keep the same order structure inherited from the assembly space. (d) Representation of the assembly space of all molecules, this space can be constrained with different structural requirements, like the type of atoms or bonds present in a molecule. More sophisticated constraints can also be imposed like, ring presence, ring size, and structural motifs.

In Assembly Theory, objects are built through recursive steps, some of the intermediate objects used to build more complicated objects may not be stable at the time of observation. These intermediate objects are referred to as virtual objects and they are not accounted for in enumerated databases with chemical filters, such as the GDB databases[6]; i.e. virtual objects are structures on assembly paths that have causal status but need not be independently observable

Although we only account for observed object types, we can still use the ordering from the Assembly space to measure the growth of chemical space with respect to complexity, Figure 1c. Therefore, in this view chemical space is a fast-growing space of unique object types with respect to complexity, Figure 1d. This space can be constrained with constraints such as the types of bonds and atoms present in a molecule or ring presence, ring size, and structural motifs. Note that the growth of these sub-spaces may be of smaller order than full chemical space, Figure 1b,d. We explore these different growth rates below.

To obtain these results, we develop a method that relies only on the number of bonds of a distribution of molecules and their atom and bond types. We use the GDB-13 database[14] as our baseline proxy for chemical space; this database enumerates a space of almost 1 billion drug-like molecules, using stability and synthetically accessible constraints on organic molecules. Using this method and the GDB-13 database, we can bound the number of molecules up to a given Assembly Index. We validate these bounds by intensive computation of the Assembly Index from the $9 \times 10^8$ molecules in the GDB-13 database. Moreover, we can constrain these bounds and obtain estimates on the size of chemical space given atom and bond type, number and size of rings in a molecule, and specific chemical motifs. Thus, this method quantifies the effort to perform an exhaustive search in varied regions of chemical space at a given level of complexity.

**Assembly Theory**

To analyse the nature of chemical space using Assembly Theory, we first define what we consider as objects and how we construct them. We consider a molecular graph as our *object* $O$, and we construct molecules by bond-joining, i.e., by joining molecular graphs at a set of vertices that are made identical[22]. Two properties of our object are of special interest for this work: the first is the *size* of the object $S(O)$, i.e. how many bonds are present in our molecular graph. The second is the *building blocks* $BB(O)$, these are the molecular graphs of size one that contain the atoms and bonds from the original object.

In Assembly Theory the *Assembly Index* $a(O)$ of an object $O$ is defined as the minimum number of recursive causal steps required to build the molecule by bond-joining operations. An *Assembly Space* is defined as a hierarchical structure in which every element is built by joining two elements in order to create a set of objects in a minimal number of steps[25]. Note that two quite different objects (e.g. in our case molecules) could have the same Assembly Index. We specify further the state of an object in Assembly Theory by combining its Assembly Index $a$ and its size $S$, as an *assembly state* $(a, S)$. We extend this idea with the following definition,

**Definition 1:** The *Assembly State Space* of an ensemble of objects is defined as the set of the assembly states of its objects.

An example of such a space is presented in Figure 2. Given that for any object $O$, its size $S$ and Assembly Index $a$ satisfy the following inequalities

$$log_2(S) \leq a \leq S - 1, \tag{1}$$

these imply that the assembly state space of any finite ensemble of objects is always bounded (e.g. the red area in Figure 2).

The right term in (1) comes from the fact that one can always build any object by joining one building block at a time. On the other hand, the left term comes from the fact that objects with the lowest number of joining operations are constructed using recursive duplication (i.e. joining the last constructed element with itself). The objects whose minimum construction corresponds to any of these previous procedures are characterized because the corresponding equality in (1) is satisfied. The objects attaining the left equality in (1) are called *duplicate-symmetric* objects, while those attaining the right equality are called *duplicate-asymmetric* objects. A subset of objects of an ensemble corresponds to a subspace in the assembly state space of the ensemble, such that its distribution reveals structural properties, Figure 2. For example, if a distribution lies more heavily on the logarithmic side it is because most of the objects are very duplicate-symmetric, while if it leans towards the linear side, it is because most of the objects are very duplicate-asymmetric.

Note that even though a molecular graph is technically a Coloured Connected Graph (CCG) (i.e. vertex- and edge-labelled connected graphs such that the vertices and edges correspond to atoms and bonds, respectively), it is constrained by several physical stability requirements (valence, ring strains, unsaturation filters, etc). In Assembly Theory[18], *Assembly Universe* refers to all possible objects with a given formal structure and all possible pathways to construct it. On the other hand, *Assembly Possible* refers to the space of physically possible objects, which are generated by the combinatorial expansion of all the known physical rules of object construction, and allows all rules to be available at every step to every object. Hence, in the terminology of Assembly Theory, the space of CCG constitutes the Assembly Universe of molecules, while the chemical space corresponds to a subspace of the Assembly Possible of molecules, Figure 1c.

Although the bounds derived below are valid for the Assembly Universe of CCG, by considering the enumerated molecules from the GDB-13 database (which is based on physical constraints and is considered as a filtered proxy of the chemical space within the Assembly Possible), we implicitly consider the nature of the physical objects and thereby obtain an estimate of the size of chemical space. Next, we derive these bounds and apply them to both the assembly universe and the chemical space of molecules.

### Estimating the Maximum and Minimum Possible Assembly Index

The bounds obtained in (1) are valid for any object in any type of ensemble. However, for the specific case of molecular graphs, we can obtain better bounds as shown in Figure 2. The possibility of computing a better upper bound for the Assembly Index than $S-1$ was pointed out qualitatively by other authors[25,26], however their results remained conjectures. Recently, bounds[27] that only depend on the size $S$ and the building blocks $BB$ have been formally derived for different types of objects, including molecular graphs, strings and polyominoes. If we consider a molecular graph $O$, then we have the following bounds[27]

$$log_2(S) \leq \ell(S) \leq a(O) \leq a_{max}(S, BB) \leq S-1, \quad (2)$$

where $\ell(S)$ is the optimal addition chain length of the size $S$ and $a_{max}$ is given by

$$a_{max}(S, BB) = \min\left\{\left\lfloor\frac{S}{2}\right\rfloor, \binom{|BB|}{2}\right\} + \left\lceil\frac{S}{2}\right\rceil - 1. \quad (3)$$

where $|BB|$ is the total number of building blocks.

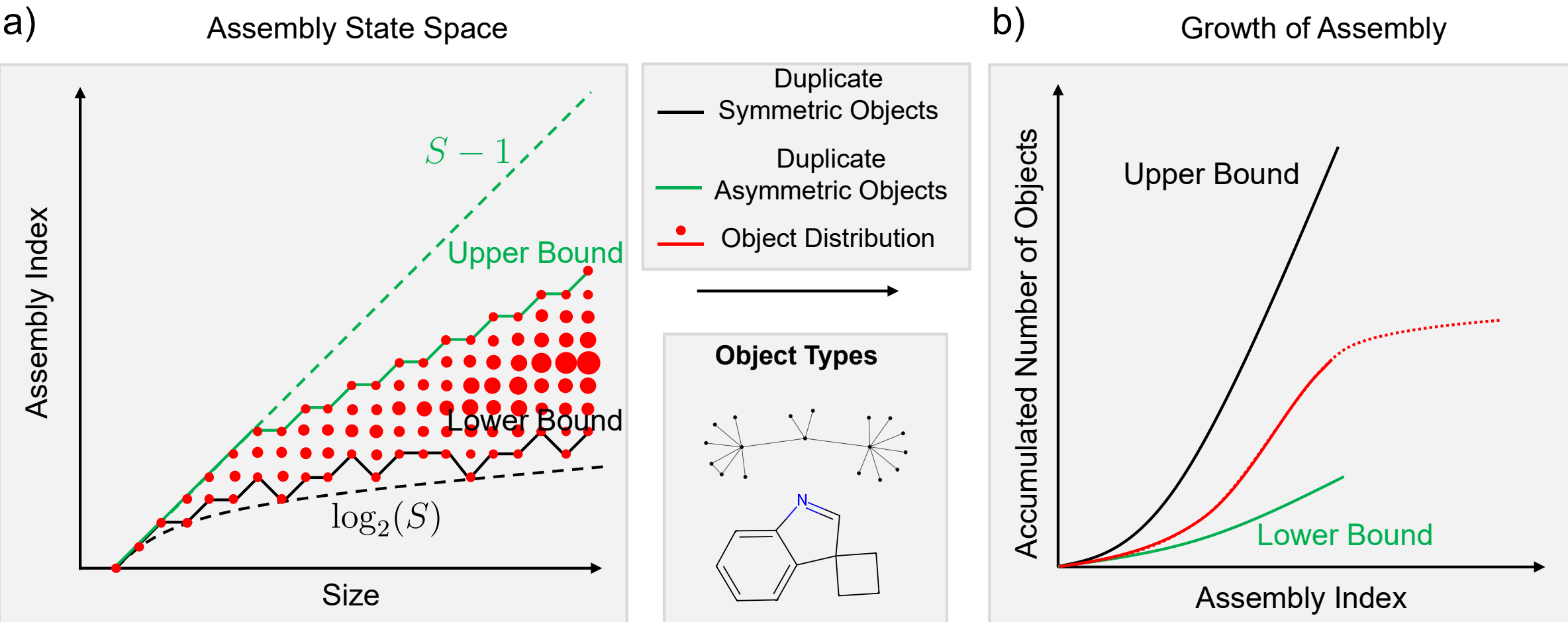

**Figure 2. Bounds on Assembly Index and the Growth of Assembly Space.** Given a distribution of objects in the assembly state space of Assembly Index and size, the distribution can be bounded given knowledge of the building blocks. The state space bounds, in conjunction with the size distribution, provide a bound on the rate of growth of the assembly space, i.e. number of objects per Assembly Index that can be constructed. The accumulated number of objects coming from finite data stops growing after the data for high Assembly Index is too sparse to explore chemical space further.

The improved lower bound in (2) comes from the fact that one can always construct a *molecular chain* using only one type of building block, while the improved upper bound is a consequence of the graph-theoretical result that any connected graph with an even number of edges *2n* can be decomposed into *n* disjoint subgraphs consisting of two edges[27]. Note that any object with an Assembly Index $a$ less than or equal to $n$ also has a $l(S)$ less than or equal to $n$. Similarly, any object with $a_{max}$ less than or equal to $n$ also has an Assembly Index $a$ less than or equal to $n$. This implies the following set inclusions

$$Set(a_{max} \leq n) \subseteq Set(a \leq n) \subseteq Set(l(S) \leq n), \tag{4}$$

and consequently, the following inequalities

$$|Set(a_{max} \leq n)| \leq |Set(a \leq n)| \leq |Set(l(S) \leq n)|, \tag{5}$$

where $|\cdot|$denotes the size of the corresponding set.

By using (5) we can estimate the size of chemical space with an Assembly Index $a \leq n$ by counting how many molecules have $a_{max}$ and $\ell(S)$ less than or equal to $n$, see Figure 2. Since the bounds only depend on size $S$ and the building blocks $BB$, we only need to know (*i*) the molecules in the chemical space considered, (*ii*) their sizes $S$ and (*iii*) the *number* of building blocks $|BB|$ used to construct it. The detailed calculations of (5) for the assembly universe and assembly possible of molecules analysed in this paper are found in the Section 3 and 4 of the Supplementary Information.

Additionally notice that even though (5) gives us a bound on the total *accumulated* number of molecules up to Assembly Index $a$ instead of a bound on the total number of molecules at a given Assembly index level. We can derive the approximate growth rates on the total number of molecules if the accumulated value up to Assembly Index $a$ is significantly larger than the accumulated value up to Assembly Index $a - 1$. For spaces growing at an exponential or higher rate and large enough $a$, this assumption holds. Finally, since the GDB-13 database would be used as numerical data to obtain the growth rates(red line), Figure 2, the data does not cover the full extent of chemical space, therefore at some level of Assembly Index the accumulated number of objects stops growing, solid to dotted line in Figure 2.

**The Rate of Growth of the Molecular Assembly Universe**

We start by estimating the size of the molecular Assembly Universe by considering the (sub)space of monochromatic Simple Connected Graphs (SCG). Even though monochromatic graphs are not the full assembly universe, they are chosen since they are well-studied[28] and exact enumeration of these is already available. Clearly, saturated hydrocarbons (alkanes) are a subspace of this universe, and these are precisely identified with the objects with one type of building block in SCG, Figure 3a. First, we estimate the size of this assembly universe and in the next section we focus on the Assembly Possible with molecules that are unsaturated hydrocarbons containing heteroatoms. The total number of SCG given a fixed number of edges is given by the sequence A002905 in the OEIS database[28], Figure 3b. From this graph we can observe that the accumulated growth of graphs with respect to the number of edges is super-exponential. With the data provided by the sequence A002905 we can calculate lower and upper bounds for the growth of the assembly space of SCG by using (5), see Figure 3c. The lower bound presents an *exponential* growth, and the upper bound a *double-exponential* growth. The bounds are computed exactly until $S = 32$ (i.e., up to $a = 5$). After this the data is too sparse to provide an accurate estimate. Next, we proceed to test the validity of these bounds by counting all SCG up to $S = 18$ edges using the `nauty` library[29] and then calculating the exact Assembly Index using an algorithm[22] specialized for large monochromatic graphs, see Figure 3c. We see from the trend that the space grows very close to the upper double-exponential bound. This supports the conjecture from previous Assembly Theory work[18] that the molecular Assembly Universe grows double-exponentially. Moreover, we see from this graph as well that most of the graphs are very duplicate-symmetric; this trend is also visible in the assembly state space of these graphs where the average distribution lies close to the symmetric bound, see Supplementary Information Section 3.1. Next, we perform this analysis for the assembly possible with the GDB-13 database as the basis of our chemical space.

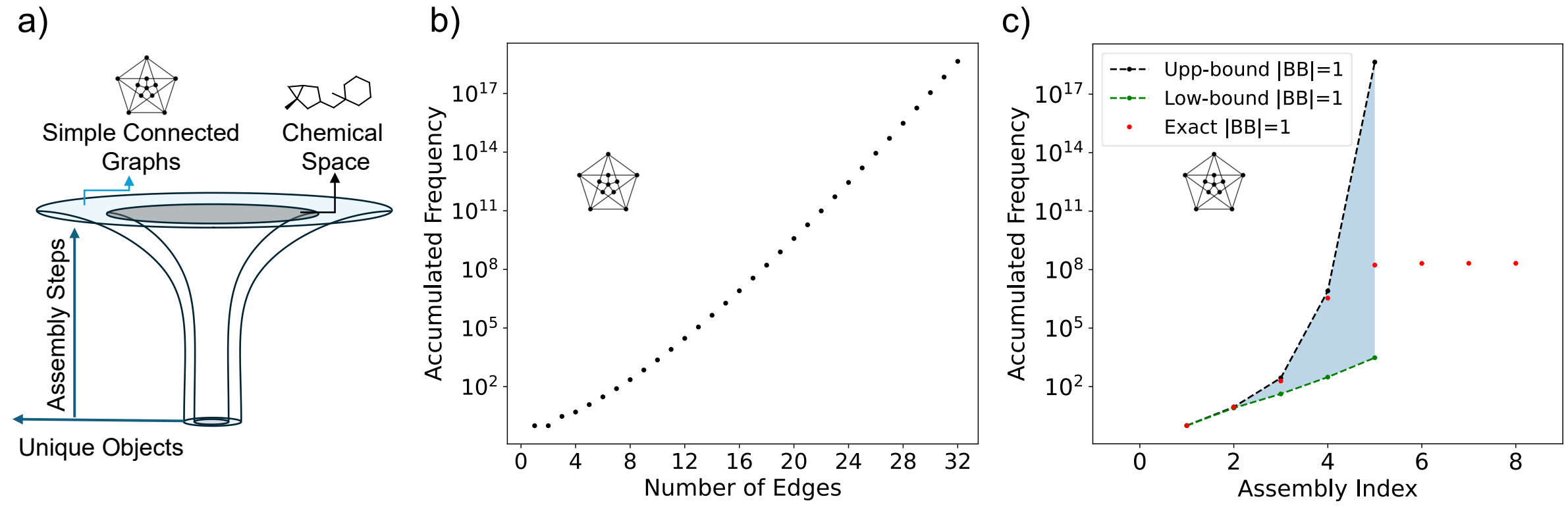

**Figure 3. Growth of the Assembly Universe of molecules.** (a) The space of all Simple Connected Graphs (SCG) for a given Assembly Index represents the assembly universe of molecules, and chemical space is a subspace of this universe representing the graphs constrained by physical construction rules. (b) Super-exponential growth of the accumulated number of SCG with a given number of edges. (c) The growth of the accumulated number of SCG given a fixed Assembly Index. The exact Assembly Index of all SCG(one type of building-block) for up to 18 edges was computed, and the bounds were computed for up to 32 edges.

## The Rate of Growth of Chemical Space

To study the geometric properties of chemical space we used the GDB-13 database[14]. It contains all small organic molecules containing up to 13 atoms of C, N, O, S and Cl. These molecules represent a subspace of the Assembly Possible since physical constraints are imposed, Figure 4a. These constraints introduce filters over the molecular Assembly Universe, e.g. removing ring strains, and add bonding and heteroatoms with unsaturation and heteroatom filters, such that the database comprises 977.358.087 molecules. The maximum number of bonds in a molecule of the GDB-13 database is $S = 19$ and the set of building blocks used to construct the molecules of this database consists of $|BB| = 19$ atomic composition-bonding type combinations.

After determining the number of molecules in the GDB-13 database for every possible number of bonds (i.e. up to 19), we can proceed to find bounds for the growth of chemical space by using (5), see Figure 4c. Since the database includes only molecules up to size 19 bonds, we have the exact behaviour of the bounds until Assembly Index 4 and incomplete afterwards, i.e. the bounds are only shown until $a = 5$ since after this point the database is too incomplete to provide an estimate. From these bounds we show that chemical space grows at *least* super-exponentially and at *most* double-exponentially with respect to the Assembly Index (at least up to $a = 4$), providing the first numerical confirmation of the conjecture that the molecular Assembly Possible grows super-exponentially[18].

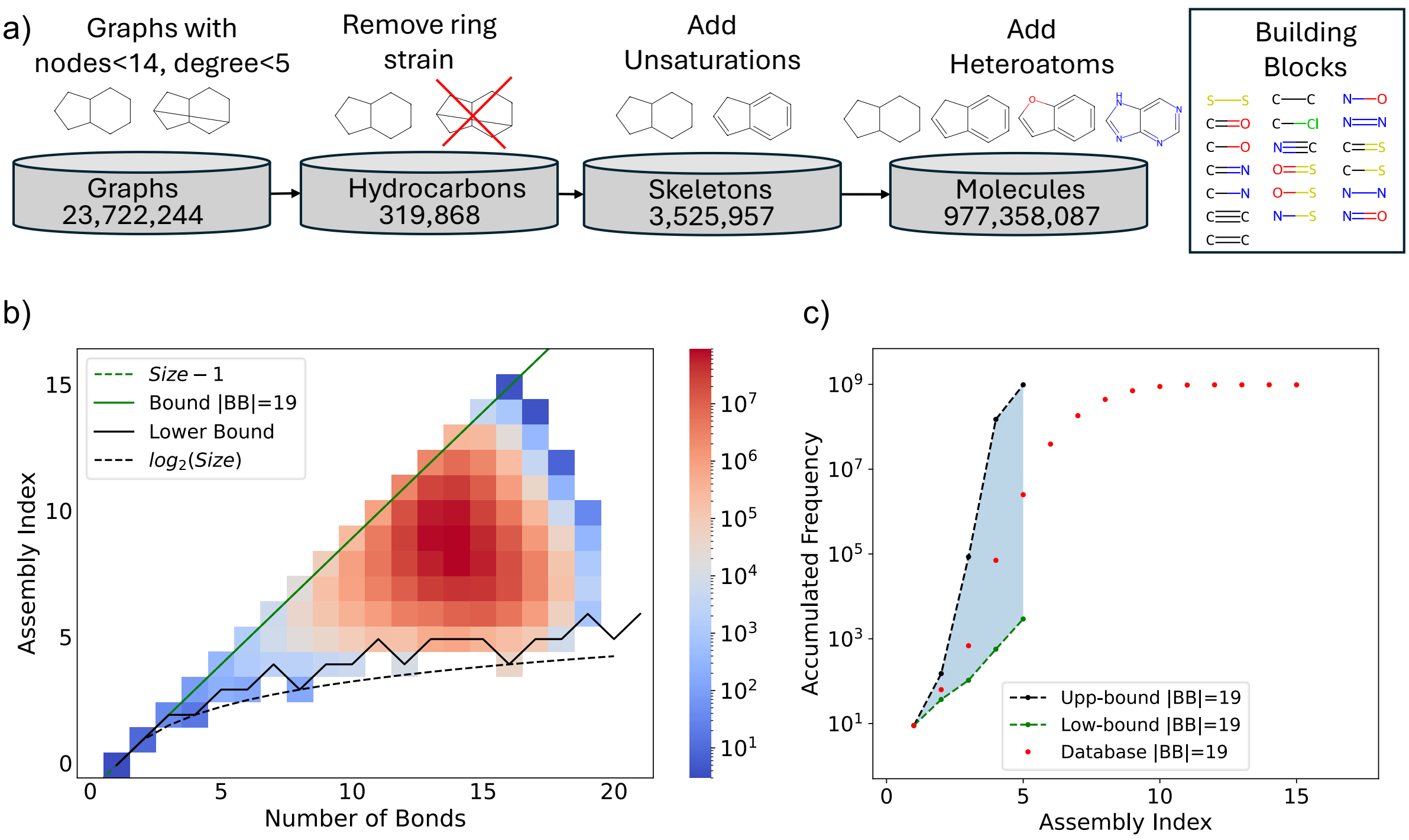

**Figure 4. Growth of the Assembly Possible proxy (chemical space) of molecules.** (a) The GDB-13 database molecule enumeration introduces a series of filters that model the physical constraints for a molecule to be possible. On the right the 19 building blocks present on the database. (b) Assembly state space of all GDB-13 molecules, the distribution is bounded by an upper bound given by the total number of bonds minus one and a linear chain lower bound. (c) The growth of the accumulated number of GDB-13 molecules with Assembly Index, the exact Assembly Index given the total number of graphs was computed for all molecules, and the bounds were computed using only the distribution of the number of molecule edges from the database.

The exact Assembly Index of all the molecules in the GDB-13 database was computed to compare with the bounds and to analyze different properties of chemical space, such as structural and motif constraints, which we explore below. We first see that the molecules are roughly evenly distributed on a logarithmic scale of the assembly state space, Figure 4b. The distribution on this scale lies between the two bounds given by (2). This is also reflected in the way the accumulated number of molecules per Assembly Index is distributed, Figure 4c. The chemical space grows at a rate between the two bounds. As we will see in the next section, this is a consequence of having both ($i$) many duplicate-symmetric molecules with few types of bonds and ($ii$) many duplicate-asymmetric molecules with lots of types of bonds and atoms. Furthermore, from this plot we have numerical evidence that the assembly possible grows super-exponentially[18].

**Constraints on Assembly Possible**

Different constraints can be introduced in chemical space, e.g., ring structure, bond length/angle, stereochemistry, chemical motifs, among others, allowing us to define and explore different regions of the space. In this work we focus on the constraints that can affect the values of our bounds (5), i.e. constraints involving a reduced number of building blocks $|BB|$ (less than the full 19 available) and modifications to the distribution of the assembly state space (Figure 4b). We start by introducing constraints on the number of building blocks $|BB|$ by considering molecules with only one, two or three types of bonds, Figures 5 and 6. For molecules with only one type of bond we consider molecules with only carbon atoms forming exclusively single bonds. This is the case where our upper bound (2) is the tightest; Figure 5a. For molecules with only two types of bond, we consider molecules with only carbon atoms forming only single or double bonds; in this case the bound is less tight since the subregion of the Assembly State Space of the GDB-13 chemical space corresponding to these type of molecules does not reach the upper boundary given by (2), Figure 5b. This is because molecules with more than $S = 12$ bonds are not exhaustively enumerated and due to the physical constraints of chemical space. As another case, we consider molecules with carbon and nitrogen atoms forming only carbon-carbon and carbon-nitrogen single bonds; in this case the upper bound (2) is reached by the Assembly State Space, Figure 5c. This is the case given that adding heteroatoms to saturated hydrocarbons explores a larger space than the physical constraints on the unsaturation of hydrocarbons. Finally, for three types of bonds we consider molecules with carbon and nitrogen atoms displaying only carbon-carbon, carbon-nitrogen single bonds and carbon-carbon double bonds. In this case the theoretical upper bound (2) is less tight since this region of the molecular Assembly Universe cannot be exhaustively explored using the molecules from GDB-13, Figure 5d.

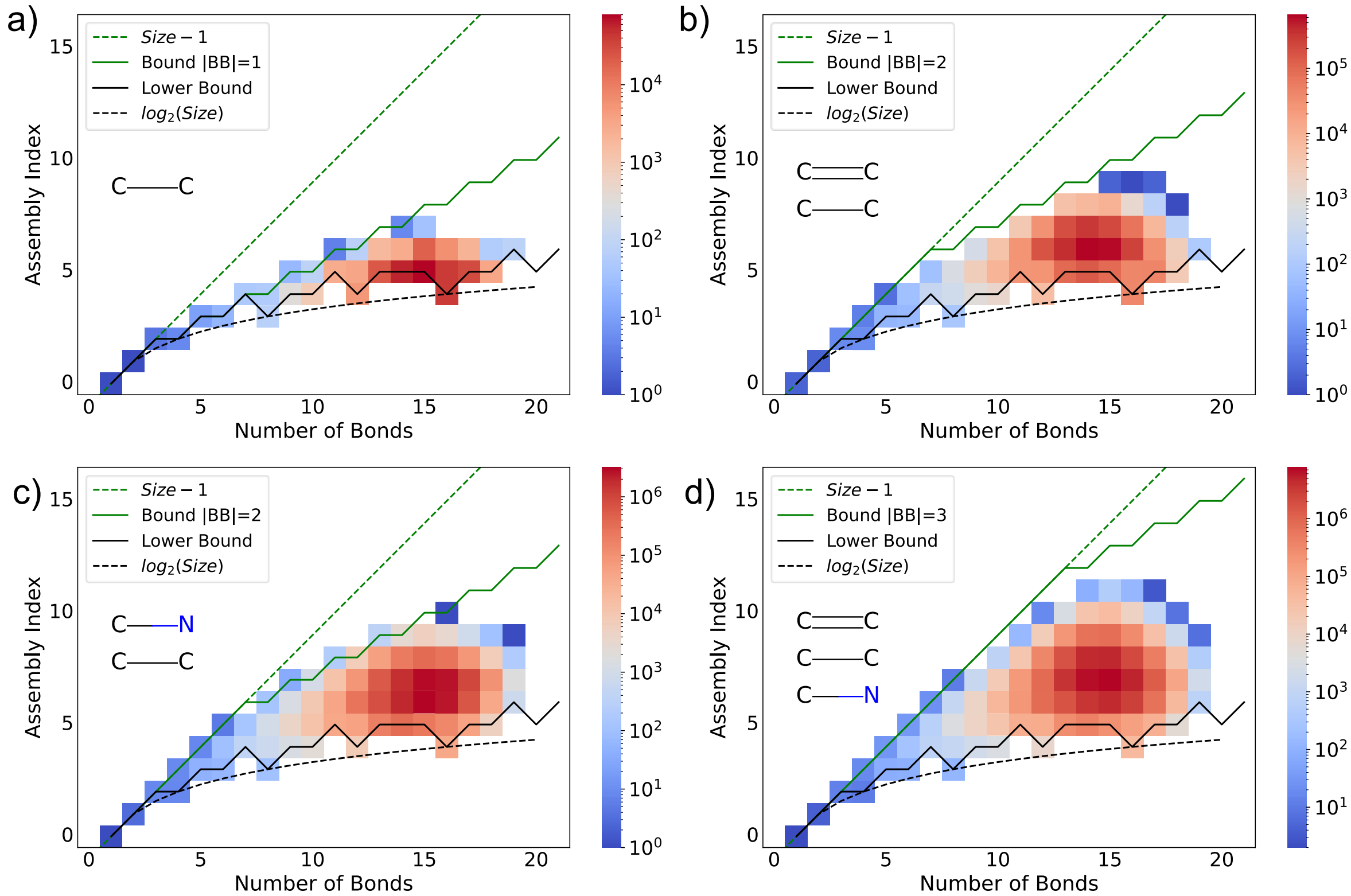

**Figure 5. Assembly State Space of Molecules with Bond Constraints.** (a) Assembly state space of all GDB-13 molecules with carbon atoms and single bonds. The distribution is bounded by a one-building block bound and a linear chain bound. (b,c) Assembly State Space of all GDB-13 molecules containing (b) carbon atoms with single, double bonds, and (c) molecules with carbon-carbon and carbon-nitrogen single bonds. The distribution is bounded by a two-building block bound and a linear chain bound. (d) Assembly state space of all GDB-13 molecules with carbon-carbon, carbon-nitrogen single bonds, and carbon-carbon double bonds. The distributions are bounded by a three-building block bound and a linear chain bound.

As in the case of the molecular Assembly Universe (Figure 3c), we find that the growth of the Assembly Space of molecules with only carbon-carbon single bonds lies close to the double-exponential, Figure 6a. We also see that the growth of chemical space is no longer so close to the double-exponential bound for both two and three types of bonds, Figure 6b,c,d. This occurs because complexity now also arises from combinations of bonds. Such complexity is of a different nature from that coming from the topological structure of the molecule. This matters because the distribution is still very close to the symmetric objects bound; therefore, the majority of molecules are still symmetric despite the combinatorial contribution of the bonds. This is a very different behaviour than in the assembly state space of strings, where the complexity comes only from the combination of the bond types, see Supplementary Information Section 3.2.

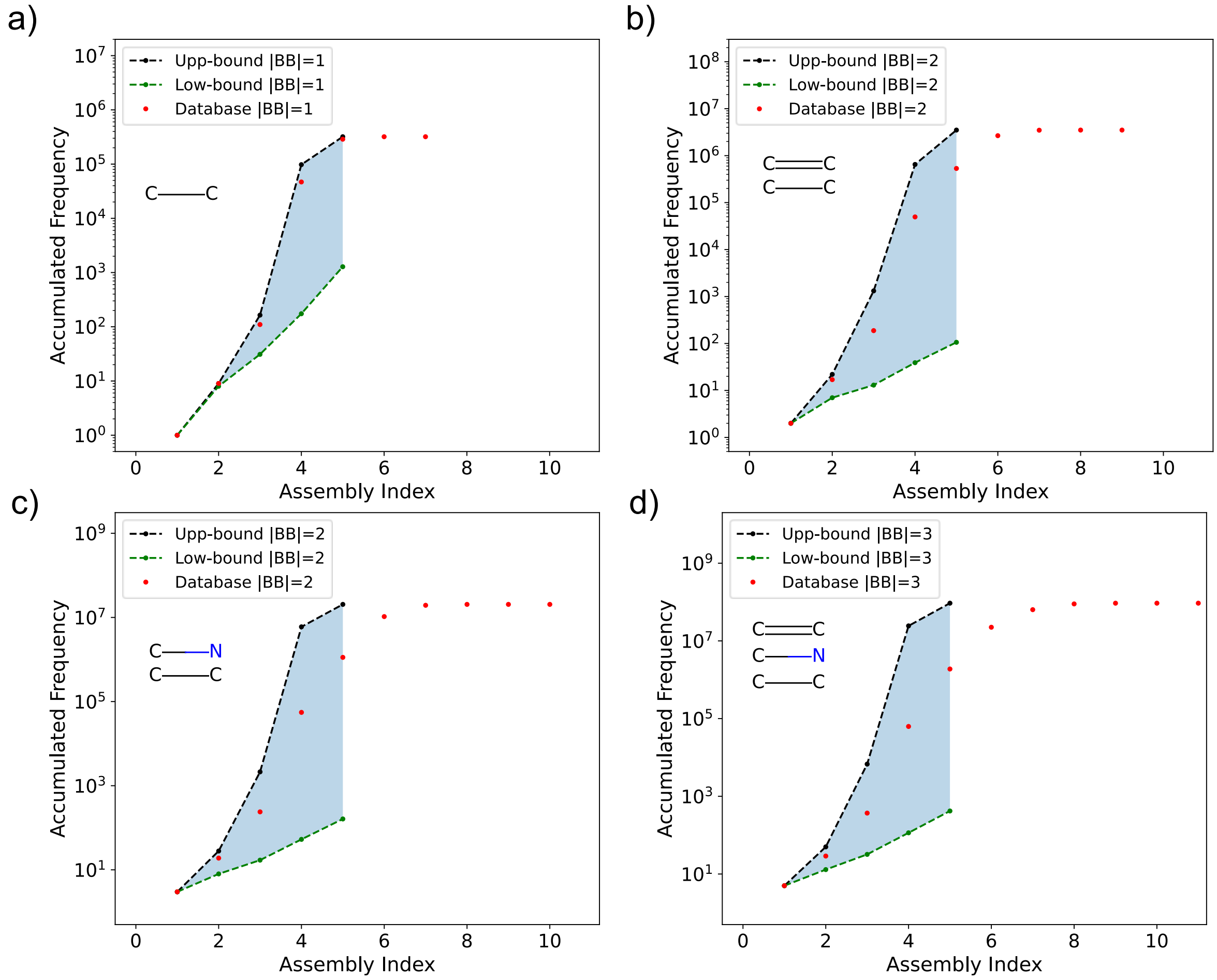

**Figure 6. Growth of chemical space with Bond Constraints.** (a) The growth of the accumulated number of GDB-13 molecules containing carbon atoms and single bonds, and their Assembly Index. (b,c) The growth of the accumulated number of GDB-13 molecules containing (b) carbon atoms with single, double bonds, and (c) carbon-carbon and carbon-nitrogen single bonds, and their Assembly Index. (d) The growth of the accumulated number of GDB-13 molecules containing carbon-carbon, carbon-nitrogen single bonds, and carbon-carbon double bonds, and their Assembly Index. The exact Assembly Index given the total number of graphs was computed for all molecules, and the bounds were computed using only the number of building blocks and the distribution of the number of molecule edges from the database.

Next, we introduce structural constraints that affect only the bond distribution of the molecules sampled. In this case, we use the number of building blocks $|BB| = 19$ that are possible in the full GDB-13 set. The bond distribution will vary by the constraints of the number of rings, the size of the ring and chemical motifs, Figure 7. Here, we vary the bond distribution by introducing several constraints, such as molecules with no rings, or with only rings, Figure 7a and 7b. In addition, we consider more specific constraints like molecules with a ring of size four, and molecules that contain cyclobutane, Figures 7c and 7d. Even though $|BB| = 19$ is the same for all these distributions, the bounds (5) are different since the distributions have different sizes because of the constraints considered, Figure 8.

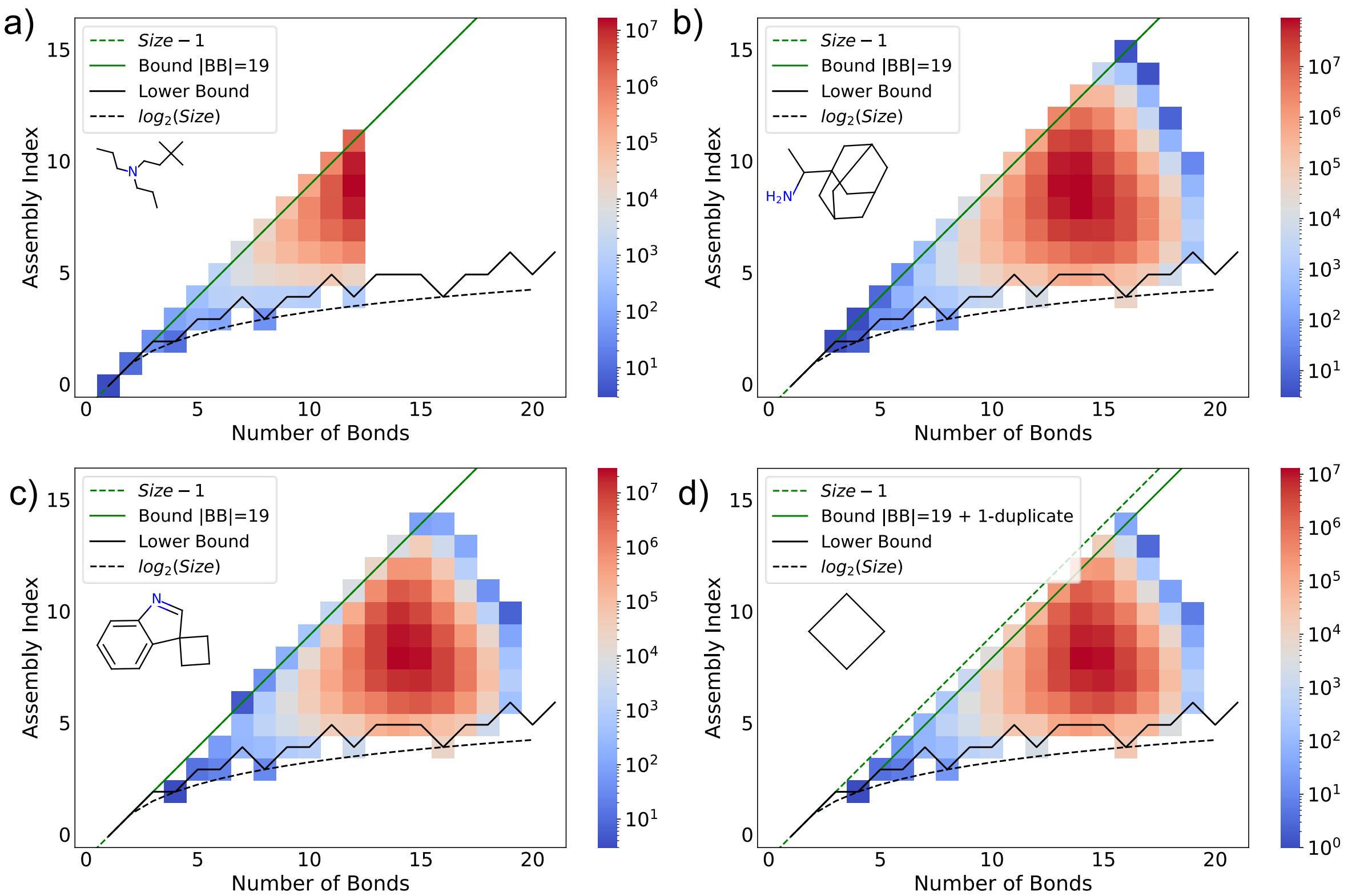

**Figure 7. Assembly State Space of Molecules with Structural Constraints.** (a,b,c,d) Assembly state space of all GDB-13 molecules without rings(a), containing rings(b), rings of size four(c), and containing cyclobutane(d). The distributions are bounded by a linear chain bound and a 19-building block bound(a,b,c) and a 19-building block bound plus a constraint of one duplicate(d).

The behaviour of the distribution of molecules with rings and without rings is clearly different in the growth of the spaces, Figure 8a and 8b. We see that for molecules without rings, the behaviour is much closer to the duplicate-asymmetric bound, whereas for the ringed molecules, it is still closer to the duplicate-symmetric bound. The presence of rings makes the molecule more duplicate-symmetric, while a non-ringed molecule can be thought of as several molecular chains put together, so the behaviour mimics that of the Assembly State Space of strings (see Supplementary Information Section 3.2).

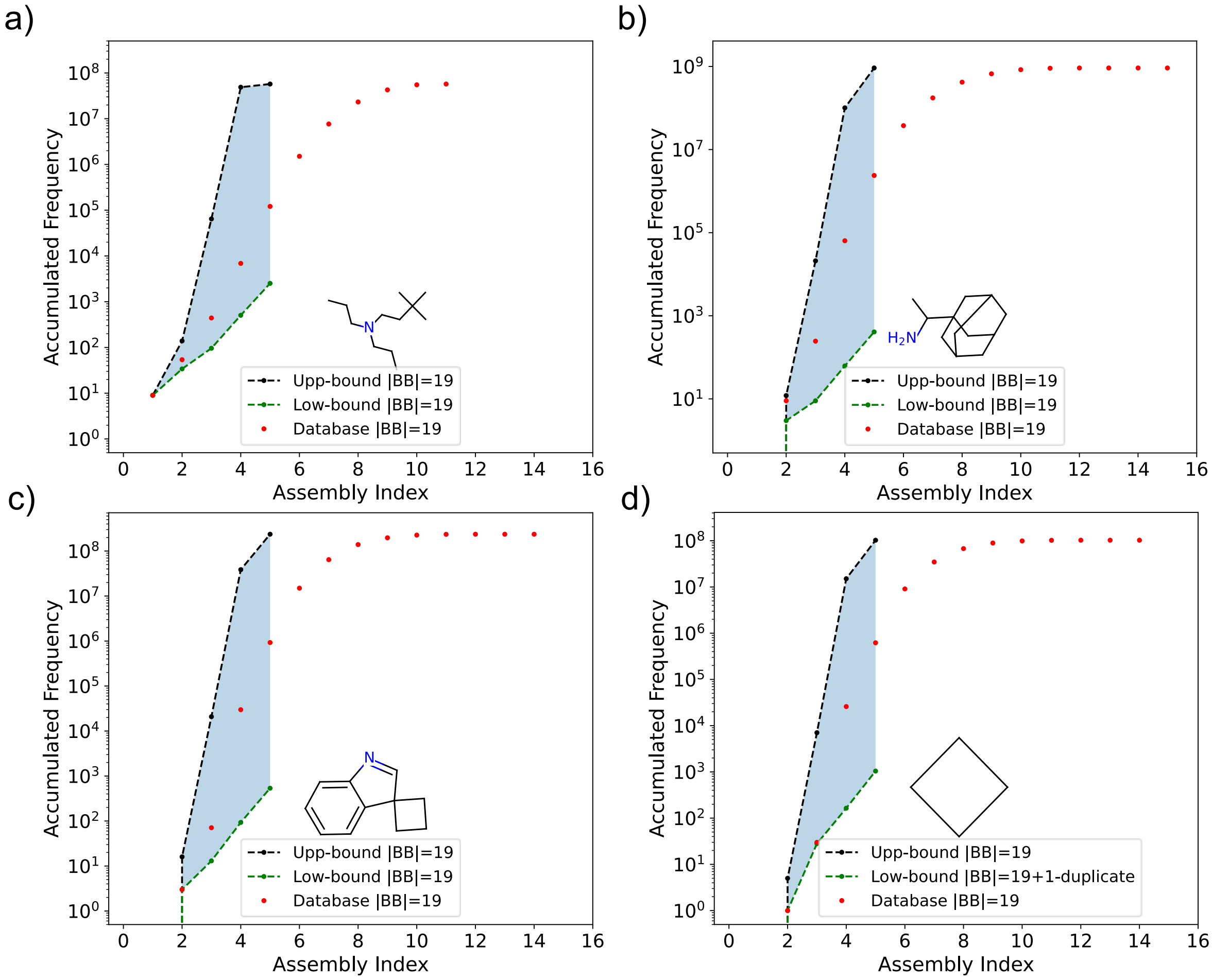

**Figure 8. Growth of chemical space with Structural Constraints.** (a,b,c,d) The growth of the accumulated number of GDB-13 molecules consisting of no rings(a), only rings(b), rings of size four(c), and containing cyclobutene(d), with respect to their Assembly Index. The exact Assembly Index given the total number of graphs was computed for all molecules, and the bounds were computed using a 19-building block bound(a,b,c) and a 19-building block bound plus a constraint of one duplicate(d), and the distribution of the number of molecule edges from the database.

We also study the growth of the chemical space of molecules constrained to contain a ring of size four, Figure 7c and 8c, and molecules that contain cyclobutene, Figure 7d and 8d. In these cases, the global behaviour is similar to that of the whole of GDB-13, with it being roughly in between the two bounds. Additionally, in the case of cyclobutene it is possible to impose an extra constraint on the bounds, since every molecule that contains it will have at least one duplicate of size two. This means that even a high molecular weight low-complexity molecule could have this minimal symmetry, see Figures 7d and 8d.

**Estimation of the Size of Chemical Space**

Next, we provide estimates of the size of the previous chemical spaces by fitting a super-exponential model to the GDB-13 data, Table 1. To see this procedure in more detail, we apply one further constraint to GDB-13 with the filters from the standard reference of the size of chemical space[1]. This includes molecules with only carbon, oxygen, nitrogen and sulfur, with an average multiplicity of 6 bonds, containing up to 4 rings and 10 branch points, Figure 9a. We fit an exponential, super-exponential and double-exponential behaviour on the lower bound, the accumulated number of computed Assembly Indexes from the database and the upper bound, Figure 9b.

| Assembly Index. | All | **Reference** | No Rings | Only Rings | 1-bond | 2-bond-a | 2-bond-b | 3-bond |
|---|---|---|---|---|---|---|---|---|
| 8 | $10^{16}$ | $\mathbf{10^{16}}$ | $10^{12}$ | $10^{14}$ | $10^{13}$ | $10^{14}$ | $10^{14}$ | $10^{15}$ |
| 10 | $10^{23}$ | $\mathbf{10^{23}}$ | $10^{17}$ | $10^{21}$ | $10^{19}$ | $10^{20}$ | $10^{21}$ | $10^{22}$ |
| 13 | $10^{37}$ | $\mathbf{10^{37}}$ | $10^{26}$ | $10^{33}$ | $10^{30}$ | $10^{32}$ | $10^{33}$ | $10^{35}$ |
| 15 | $10^{48}$ | $\mathbf{10^{48}}$ | $10^{32}$ | $10^{42}$ | $10^{38}$ | $10^{40}$ | $10^{42}$ | $10^{44}$ |
| 25 | $10^{118}$ | $\mathbf{10^{117}}$ | $10^{74}$ | $10^{101}$ | $10^{88}$ | $10^{96}$ | $10^{101}$ | $10^{107}$ |
| 30 | $10^{161}$ | $\mathbf{10^{160}}$ | $10^{99}$ | $10^{136}$ | $10^{120}$ | $10^{130}$ | $10^{136}$ | $10^{147}$ |

**Table 1. Size Estimation of chemical spaces**. Total Size of chemical spaces given different Assembly Index values obtained by a super-exponential fitting of the computed Assembly Indexes from GDB-13. All refers to the full GDB-13 database from Figure 4, Filtered to the data from Figure 9, No Rings, Only Rings the data from Figure 8a,b respectively and 1-bond, 2-bond-a, 2-bond-b and 3-bond refers to Figure 6a,b,c,d respectively.

Using the fitted super-exponential model of the reference-filtered chemical space1, we estimate the size of the space for different assembly indices, Table 1. The same procedure is applied with the full GDB-13 data and the previously studied chemical spaces with building-block and ring constraints, details can be found in the Supplementary Section 4.3. To compare to previous studies of the size of chemical space, we estimate 25 to be the maximum Assembly Index that a molecule with molecular mass of <500 Dalton would have. To do this, the highest Assembly Index molecules with molecular mass of <500 Dalton from PubChem[5] and COCONUT[30] databases were queried, and from the molecules that satisfied the reference filters, the Assembly Index 25 was identified. From this estimate we obtain a size of chemical space of $10^{117}$, which is considerably higher than the size[1] of $10^{60}$. For a proper comparison we note that the Assembly Index 25 molecules had a maximum of 36 atoms, which following the reference[1] logic yields a chemical space of $10^{70}$. Therefore, our analysis yields a much larger chemical space growth than the reference assumptions. Also of importance is the size of the space of Assembly Index 15, which yields $10^{48}$, since this is the limit that has been established in previous studies of abiotic/biotic transition[31].

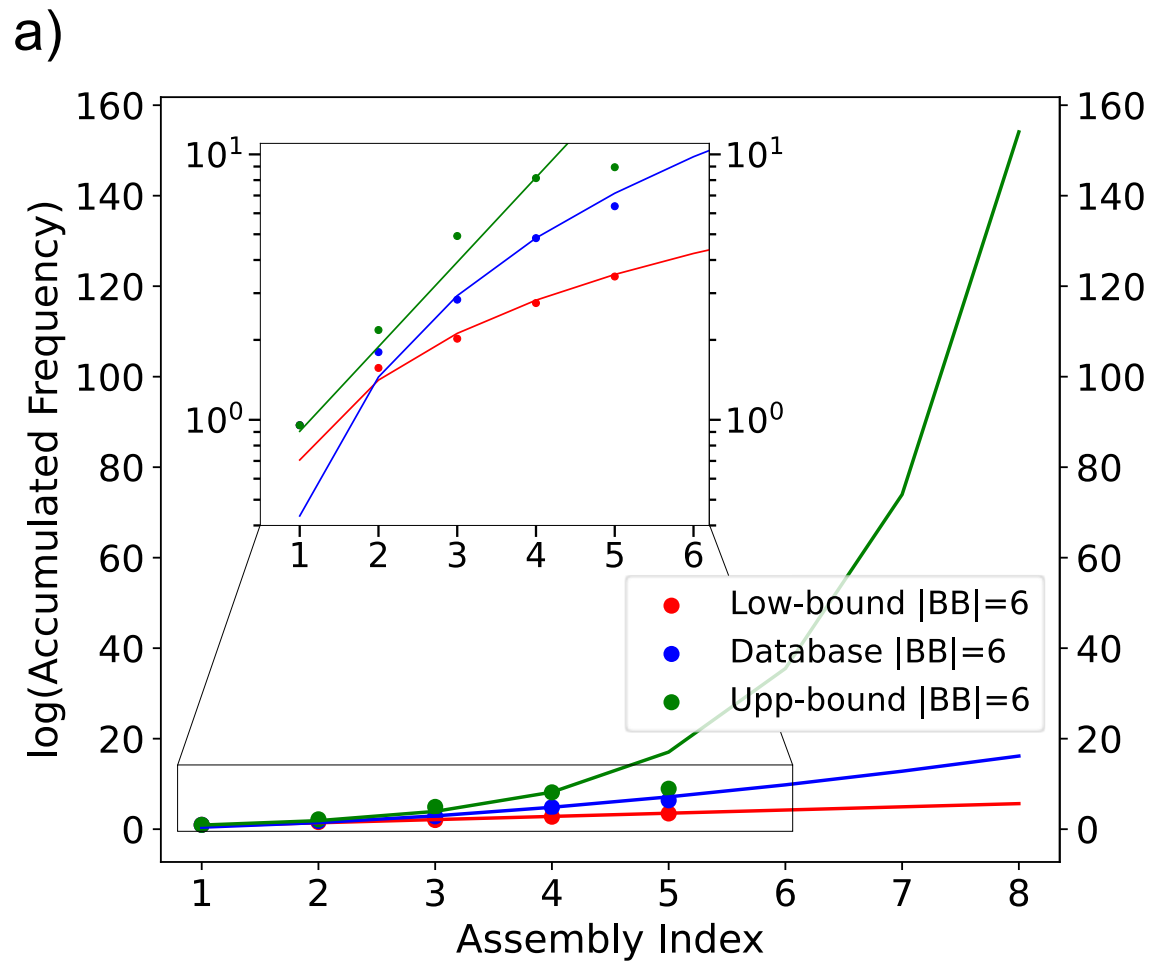

b)

| Growth | Fit | Constants | Colour |
|---|---|---|---|
| Exponential | $e^{b \cdot a}$ | $b = 1.62$ | Red |
| Super-Exponential | $e^{a^c}$ | $c = 1.73$ | Blue |
| Double-Exponential | $e^{e^{d \cdot a}}$ | $d = 0.73$ | Green |

**Figure 9. Growth of reference-filtered chemical space**. a) The growth of the accumulated number of GDB-13 molecules up to 6 building blocks and filtered by excluding Cl atoms, more than 4 rings, and 10 branches. An exponential, super-exponential and double-exponential fitting is applied to the lower bound, the computed Assembly Indexes from the database and the upper bound respectively. b) Functional forms and parameters of the fitted parameters to the lower bound, the computed Assembly Indexes from the database and the upper bound.

Finally, we performed a similar chemical space size estimation using a 50 million subsample of GDB-17 [15]. For this, we found the intersection between this subsample and GDB-13, then we used an escort scaling model to find the scaling model of this intersection to the full size of GDB-13. Using this scaling model we took the subsample of GDB-17 with the reference constraints[1] applied previously and scaled this constrained subsample. We show that with this independent scaled database we obtain a super-exponential fitting of $c = 1.71$, Figure 9b. This is close to the GDB-13 fitting, thus giving more confidence to our estimation methodology. This gives an estimate of chemical space size of $10^{107}$, at Assembly Index 25. A smaller value than the GDB-13 estimation, however since this is only an approximation of GDB-17, this value has lower certainty than the value of GDB-13. For full details see the Supplementary Section 4.4.

**Exploration of the Assembly State Space of Motifs**

To explore chemical space constrained by specific chemical motifs we selected the two common biologically important heterocycles, purine and pyrimidine, and restricted the chemical space to all molecules that contain these motifs, as shown in Figure 10. These molecules were chosen because of their biological relevance, and since their size is small enough to constrain GDB-13. In these two cases, our bounds are constrained by the motifs, since the sizes of the sets of molecules are obviously motif dependent. Across both cases, all structures will always have one duplicate of two and three bonds for pyrimidine and purine, respectively. This is shown in the upper bounds for the assembly state spaces of both molecules, Figure 10. Since for both motifs, the explored chemical space has Assembly Index greater than or equal to four, the estimates of the growth of chemical space are not accurate because the number of explored molecules with a high number of bonds is too small. The estimation of the size of chemical space given a sophisticated motif, such as a large molecule or a large polymer like RNA, is still an open problem. However, we can study the nature of these chemical subspaces by probing the boundaries of the assembly state space defined by the restriction of the motifs in the GDB-13 database. For both motifs, we sampled the boundaries of the assembly state space and filtered the molecules with a force field-based energy minimization. Molecular geometries were optimized using the Merck Molecular Force Field (MMFF94) as implemented in RDKit. Some structural patterns emerged from the combination of the constraint and the assembly process. Molecules with low Assembly Index are symmetric and often contain several overlapping rings. High Assembly Index molecules contain additional heteroatoms, bonds, and often sophisticated ring structures. For an extensive view of the sampled molecules, see Supplementary Information Section 4.5.

A more comprehensive study of motifs in the assembly state space can be performed by introducing a selection mechanism in the chemical space exploration[19,20]. This could probe the directed and undirected exploration dynamics that are characteristic of assembly spaces[18]. Here, by contrast, we mainly explored the undirected exploration dynamics present in the Assembly Universe and the Assembly Possible. Such directed exploration and selection are beyond the scope of this work.

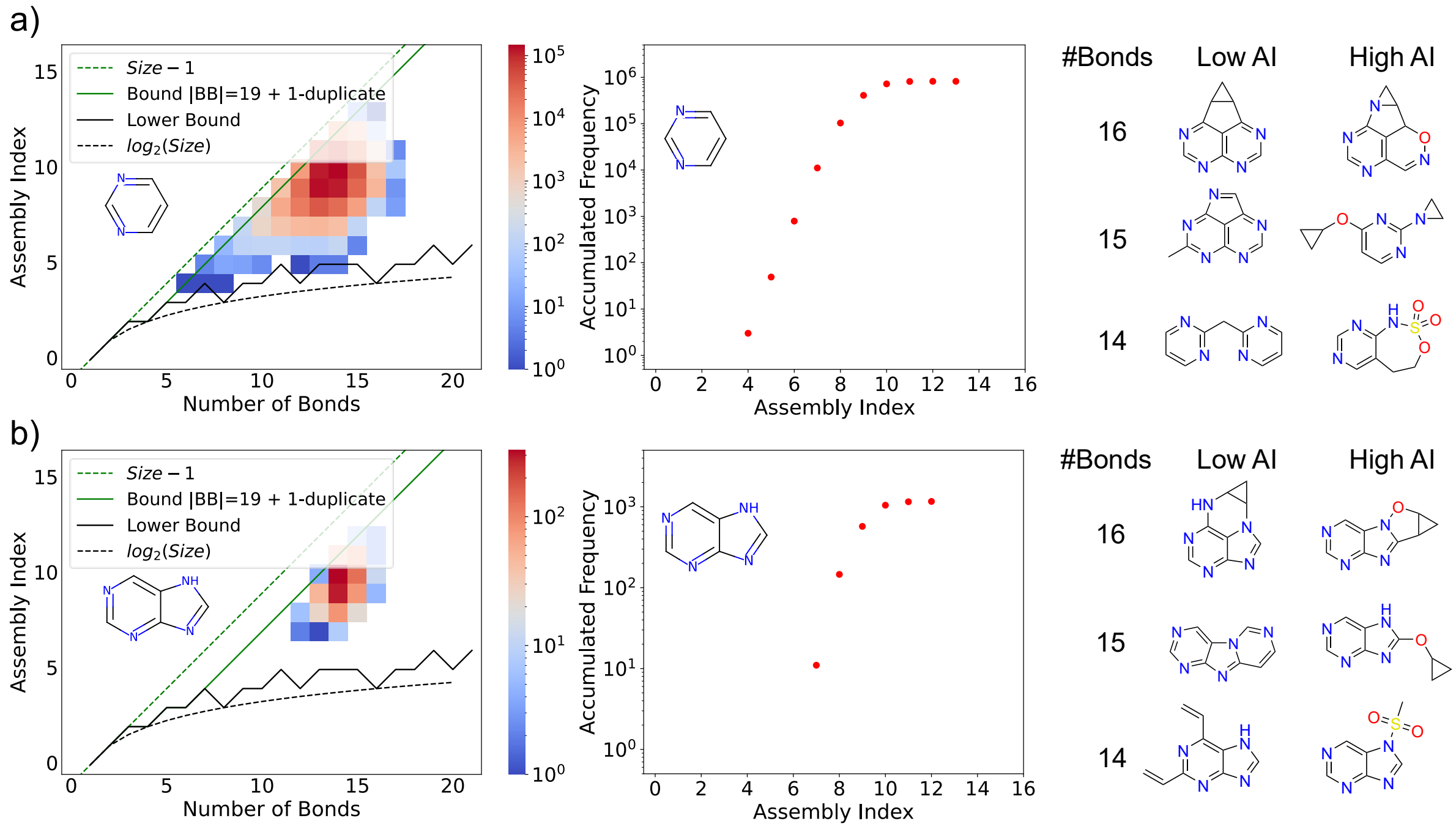

**Figure 10. Assembly Space Exploration, given a constraint of Structural Motifs.** (a,b) Assembly state space of all GDB-13 molecules containing a pyrimidine(a), a purine(b) motif, and the growth of the total accumulated number of such molecules with Assembly Index. Molecules with those constraints with both high and low assembly indices(AI) are sampled to find compelling molecules of structural relevance.

## Conclusions

In this paper we introduce a novel approach to estimate the size of chemical space using the molecular complexity measure known as the Assembly Index. The mapping of the chemical space into the Assembly State Space allows us to parametrize the former in terms of the Assembly Index and number of bonds of the molecules. Using mathematical bounds that constrain the minimum $\ell(S)$ and maximum $a_{max}(S, BB)$ values of Assembly Index that a molecule with $S$ number of bonds and $|BB|$ types of bonds could exhibit, this mapping allows us to compute upper and lower bounds for the size of the subspace of molecules in the chemical space with Assembly Index less than or equal to a fixed value $a$. The growth of these upper and lower bounds in terms of $a$ exhibits double-exponential and exponential behaviours, respectively, implying that the growth of the chemical space as a function of the Assembly Index $a$ is super-exponential. The lower and upper bounds were additionally validated for several subspaces of the chemical space using the GDB-13 database as a model for chemical space. These subspaces were molecules composed of carbon atoms with (*a*) only single bonds, (*b*) single and double bonds, molecules composed of (c) carbon-carbon, carbon-nitrogen single bonds, and (*d*) carbon-carbon, carbon-nitrogen single bonds, and carbon-carbon double bonds. Moreover, structurally constrained subspaces of the chemical space were also used to test these bounds, e.g., molecules with (*i*) a fixed number of rings, (*ii*) a fixed size of rings and (*iii*) specific chemical motifs. Finally, using the reference constraints[1], the size of the chemical space is inferred to be of the order of $10^{117}$. This is based on the estimate that the maximum Assembly Index that a molecule with molecular mass of <500 Dalton could exhibit is 25.

The methodology introduced in this paper to quantify the growth of the chemical space could be extended to estimate the growth of *any* set of objects constructible from a finite number of building blocks, that is, any Assembly Space. The mapping into the Assembly State Space provides a parametrization of any object in terms of its Assembly Index and size, such that establishing bounds over this latter space translates into lower and upper bounds for the size of the Assembly Space. This formalism was applied first to chemical space exploration by fixing natural product motifs or structural features of molecules of interest, obtaining estimates of the size of chemical space and finding compelling new molecules in the boundaries of complexity in the Assembly State Space.

## Methods

All Assembly Index calculations using the graph assembly calculator were performed with the library https://github.com/croningp/assemblycpp-v5. The data analysis was performed using Python 3. Complete methodological details are provided in the Supplementary Information.

## Code Availability

All the codes required for assembly calculations and generating the figures are available at: Before publication it is here:

## Author Contributions

L.C. conceived the concept and the research plan and explored the idea together with K.Y.P. and J.C.M.P.; K.Y.P. developed the construction for the maximum possible Assembly Index. J.C.M.P. extended the results of the maximum possible Assembly Index and used it to bound the growth of assembly spaces and impose constraints on them. D.O.A. and A.S. computed the assembly indices for GDB-13. K.Y.P. did the calculations for the estimation of the size of chemical space. K.Y.P., J.C.M.P., and L.C. wrote the manuscript.

## Acknowledgements

We are grateful to Stuart Marshall for carefully reading and providing useful comments about a preliminary draft of the present document. We acknowledge financial support from the Sloan Foundation (grant number G-2023-21110), the John Templeton Foundation (grant nos. 61184 and 62231), the Engineering and Physical Sciences Research Council (EPSRC) (grant nos. EP/L023652/1, EP/R01308X/1, EP/S019472/1 and EP/P00153X/1), the Breakthrough Prize Foundation and NASA (Agnostic Biosignatures award no. 80NSSC18K1140), MINECO (project CTQ2017-87392-P) and the European Research Council (ERC) (project 670467 SMART-POM).

## References

1. Bohacek, R. S., McMartin, C. & Guida, W. C. The art and practice of structure-based drug design: A molecular modeling perspective. *Med. Res. Rev.* **16**, 3–50 (1996).
2. Davies, D. W. *et al.* Computational Screening of All Stoichiometric Inorganic Materials. *Chem* **1**, 617–627 (2016).
3. Walsh, A. The quest for new functionality. *Nat. Chem.* **7**, 274–275 (2015).
4. Coley, C. W. Defining and Exploring Chemical Spaces. *Trends Chem.* **3**, 133–145 (2021).
5. Kim, S. *et al.* PubChem 2025 update. *Nucleic Acids Res.* **53**, D1516–D1525 (2025).
6. Reymond, J.-L. The Chemical Space Project. *Acc. Chem. Res.* **48**, 722–730 (2015).
7. Henault, E. S., Rasmussen, M. H. & Jensen, J. H. Chemical space exploration: how genetic algorithms find the needle in the haystack. *PeerJ Phys. Chem.* **2**, e11 (2020).
8. Verhellen, J. & Van Den Abeele, J. Illuminating elite patches of chemical space. *Chem. Sci.* **11**, 11485–11491 (2020).
9. Jensen, J. H. A graph-based genetic algorithm and generative model/Monte Carlo tree search for the exploration of chemical space. *Chem. Sci.* **10**, 3567–3572 (2019).
10. Skinnider, M. A., Stacey, R. G., Wishart, D. S. & Foster, L. J. Chemical language models enable navigation in sparsely populated chemical space. *Nat. Mach. Intell.* **3**, 759–770 (2021).
11. Zhang, K. *et al.* Artificial intelligence in drug development. *Nat. Med.* **31**, 45–59 (2025).

12. Bilodeau, C., Jin, W., Jaakkola, T., Barzilay, R. & Jensen, K. F. Generative models for molecular discovery: Recent advances and challenges. *WIREs Comput. Mol. Sci.* **12**, e1608 (2022).

13. Drew, K. L. M., Baiman, H., Khwaounjoo, P., Yu, B. & Reynisson, J. Size estimation of chemical space: how big is it? *J. Pharm. Pharmacol.* **64**, 490–495 (2012).

14. Blum, L. C. & Reymond, J.-L. 970 Million Druglike Small Molecules for Virtual Screening in the Chemical Universe Database GDB-13. *J. Am. Chem. Soc.* **131**, 8732–8733 (2009).

15. Ruddigkeit, L., Van Deursen, R., Blum, L. C. & Reymond, J.-L. Enumeration of 166 Billion Organic Small Molecules in the Chemical Universe Database GDB-17. *J. Chem. Inf. Model.* **52**, 2864–2875 (2012).

16. Polishchuk, P. G., Madzhidov, T. I. & Varnek, A. Estimation of the size of drug-like chemical space based on GDB-17 data. *J. Comput. Aided Mol. Des.* **27**, 675–679 (2013).

17. Buehler, Y. & Reymond, J.-L. A View on Molecular Complexity from the GDB Chemical Space. *J. Chem. Inf. Model.* **65**, 8405–8410 (2025).

18. Sharma, A. *et al.* Assembly theory explains and quantifies selection and evolution. *Nature* **622**, 321–328 (2023).

19. Liu, Y. *et al.* Exploring and mapping chemical space with molecular assembly trees. *Sci. Adv.* **7**, eabj2465 (2021).

20. Pagel, S., Sharma, A. & Cronin, L. Mapping Evolution of Molecules Across Biochemistry with Assembly Theory. Preprint at https://doi.org/10.48550/ARXIV.2409.05993 (2024).

21. Patarroyo, K. Y. *et al.* Quantifying the Complexity of Materials with Assembly Theory. Preprint at https://doi.org/10.48550/ARXIV.2502.09750 (2025).

22. Seet, I., Patarroyo, K. Y., Siebert, G., Walker, S. I. & Cronin, L. Rapid Exploration of the Assembly Chemical Space of Molecular Graphs. *J. Chem. Inf. Model.* acs.jcim.5c01964 (2025) doi:10.1021/acs.jcim.5c01964.

23. Rutter, L. A. *et al.* Exploring molecular assembly as a biosignature using mass spectrometry and machine learning. Preprint at https://doi.org/10.48550/ARXIV.2507.19057 (2025).

24. Kempes, C. P. *et al.* Assembly theory and its relationship with computational complexity. *Npj Complex.* **2**, 27 (2025).

25. Marshall, S. M., Moore, D. G., Murray, A. R. G., Walker, S. I. & Cronin, L. Formalising the Pathways to Life Using Assembly Spaces. *Entropy* **24**, 884 (2022).

26. Łukaszyk, S. & Bieniawski, W. Assembly Theory of Binary Messages. *Mathematics* **12**, 1600 (2024).

27.Cronin, L., Parra, J. C. M. & Patarroyo, K. Y. Assembly Addition Chains. Preprint at https://doi.org/10.48550/ARXIV.2512.18030 (2025).

28. OEIS Foundation Inc. The On-Line Encyclopedia of Integer Sequences. *Published electronically at https://oeis.org/A002905*.

29. McKay, B. D. & Piperno, A. Practical graph isomorphism, II. *J. Symb. Comput.* **60**, 94–112 (2014).

30. Sorokina, M., Merseburger, P., Rajan, K., Yirik, M. A. & Steinbeck, C. COCONUT online: Collection of Open Natural Products database. *J. Cheminformatics* **13**, 2 (2021).

31. Marshall, S. M. *et al.* Identifying molecules as biosignatures with assembly theory and mass spectrometry. *Nat. Commun.* **12**, 3033 (2021).